\documentclass[11pt]{article}
\usepackage{graphicx}
\usepackage[a4paper,margin=1.5cm]{geometry}
\usepackage{amssymb}
\usepackage{natbib}
\usepackage{amsmath}
\usepackage{mathtools}
\usepackage{tensor}
\usepackage{color}

\title{Zero Mode Tunnelling in a Fractional Quantum Hall Device}

\author{S. Huntington and V. Cheianov \\
\\ \vspace{4pt} Department of Physics, Lancaster University, Lancaster LA1 4YB,
United Kingdom}

\begin{document}
\maketitle

\begin{abstract}

  Tunnelling measurements on fractional quantum Hall systems are continuing to
  increase in popularity since they provide a method to probe the non-Fermi
  liquid behaviour of fractionally charged excitations occupying the edge states
  of a quantum Hall system. When considering tunnelling one must resort to an
  effective theory and typically a phenomenological tunnelling Hamiltonian is
  used analogous to that used for a conventional Luttinger liquid. It is the
  form of this tunnelling Hamiltonian that is investigated in this work by
  making a comparison to an exact microscopic calculation of the zero mode
  tunnelling matrix elements. The computation is performed using Monte Carlo and
  results were obtained for various system sizes for the $\nu=1/3$ Laughlin
  state. Here we also present a solution to overcome the phase problem
  experienced in Monte Carlo calculations using Laughlin-type
  wavefunctions. Comparing the system size dependence of the microscopic and
  phenomenological calculations for the tunnelling matrix elements, it was found
  that only for a particular type of operator ordering in the tunnelling
  Hamiltonian was it possible to make a good match to the numerical
  calculations. From the Monte Carlo data it is also clear that for any system
  size the electron tunnelling is always less relevant than the quasiparticle
  tunnelling process, supporting the idea that when considering tunnelling at a
  weak barrier, the electron tunnelling process can be neglected.

\end{abstract}


One of the key signatures of the fractional quantum Hall effect (FQHE) are
plateaus in the Hall conductance which are given by some rational factor, $\nu$
multiplied by the ratio $e^2/h$ \cite{fqhe_observation1}. It is well understood
that the bulk states of the system are incompressible \cite{laughlinPA} and thus
transport properties are completely determined by 1D channels occupying the
edges of the FQH droplet. These 1D channels consist of interacting electrons for
which the Fermi liquid theory breaks down and it was first proposed by Wen
\cite{wenChiLL} that they should instead, be described by a chiral Luttinger
liquid. A chiral Luttinger liquid has the key feature of low-energy excitations
being collective sound modes and it can be shown that the system has a
four-terminal Hall conductance given by $\nu(e^2/h)$ \cite{wenChiLL}. These
features are similar to a conventional Luttinger liquid in which there are left
and right moving particles along the same one dimensional channel. The chiral
Luttinger liquid is a phenomenological theory useful for describing transport
properties of the FQHE which are readily accessible to experimental
measurements.

More recently, a great amount of theoretical and experimental effort has focused
on the transport properties of FQH edge states when the charge carriers are
faced with a single, or multiple potential barriers and tunnelling is observed. 
There are two equivalent methods to initiate tunnelling in a QH device. The most
common from an experimentalists point of view is by physically moving the edges
closer together at some point along a Hall bar, known as a quantum point contact
(QPC). Such a constriction can be achieved by placing metallic plates above the
2D electron gas and applying a negative bias causing a local depletion of
electrons. As a result edge states are brought closer together causing a finite
probability of inter edge back scattering. The strength of this pinching effect
on the edge states is determined by the magnitude of the bias applied to the
magnetic plates. For ideal systems, the same tunnelling behaviour can be obtained
from placing an impurity into the bulk which also couples the edges and allows
back scattering to occur. Realistically using an impurity is a much cleaner
method to observe basckscattering since a QPC can have adverse effects on the
surrounding quantum Hall fluid due to electrostatic
reconstruction as discussed in the review by Chang \cite{RevModPhys.75.1449}.

One of the first pieces of work concerned with tunnelling at a QPC was carried
out by Kane and Fisher
\cite{kane1992transmission,PhysRevB.46.7268,PhysRevLett.68.1220} who
investigated tunnelling at both a weak link and a weak barrier in a conventional
Luttinger liquid. Similar work was also carried out by Furusaki and Nagaosa
\cite{PhysRevB.47.4631} and Moon \textit{et al.}
\cite{PhysRevLett.71.4381}. At a weak link tunnelling will be dominated by
electrons in the Luttinger liquid since, effectively the liquid is split into
two separate islands. For a weak barrier however it is the excitations of the
Luttinger liquid that tunnel. For the work carried out by Kane and Fisher a
perturbative approach for the tunnelling Hamiltonian was used in conjunction
with RG to discover which of the tunnelling processes were relevant in both the
strong and weak back scattering limit and predictions were made about the
tunnelling conductance and the zero-bias peaks in I-V characteristics. The
predictions about the Luttinger liquid behaviour show stark contrasts to that of
the non-interacting system, the Fermi liquid. Here, unlike in the
non-interacting case, the width of the zero-bias peaks are temperature
dependent, and in particular the conductance away from the peak has a power law
temperature dependence where the exponent of the power law is the interaction
parameter of the Luttinger liquid. The reason for the differences is of course
down to the interactions between the Fermions in 1D and thus observing such
behaviour in a 1D channel will be key in discovering systems that are strongly
correlated, non-Fermi liquids. The form of the tunnelling Hamiltonian used in
all of the referenced works in this report can be mapped onto the boundary
sine-Gordon model \cite{gogolin2004bosonization} and consists of operators that
annihilate a charge carrier in one direction and create another carrier
traveling in the opposite direction at some barrier.

So far there is experimental agreement of non-Fermi liquid behaviour in the
Laughlin type edge states of a FQH device
\cite{Milliken1996309,PhysRevLett.93.046801} though the specific value of the
power of the temperature dependence of the tunnelling conductance is slightly
off the expected theoretical value \cite{PhysRevLett.77.2538}. Experiments
measuring shot noise and interference experiments (all making use of one or
multiple point contacts) are predicted to prove the existence of fractionally
charged carriers in the edge states as well as display their (Abelian or
non-Abelian) fractional statistics. In particular it has been predicted that for
Laughlin type QH states, the back scattered current in shot noise experiments
should be proportional to the charge of the carriers \cite{PhysRevLett.72.724}
given by $e^* = \nu e$ in the weak back scattering limit at zero temperature,
where $\nu$ is the filling fraction of the lowest Landau level. Experimental
work has claimed to have observed Laughlin type quasiparticles in such
experiments \cite{shotnoise_support2,shotnoise_support1} though not all of the
work agrees that the shot noise measurement is dependent only on the charge of
the quasiparticles. It has been claimed that the specifics of the tunnelling
barrier as well as the energy regimes used in the experiment can effect the
value of the back scattered current, this would account for a deviance in the
predicted value of the quasiparticle charge in the very weak back scattering
limit obtained in some experiments \cite{shotnoise_question}. In particular, the
boundary sine-Gordon model for various test states has not been able to resolve
which ground state provides a good description for even denominator filling
fractions such as $\nu=5/2$. Experiments at quantum point contacts for the
$\nu=5/2$ states which measure tunnelling noise and tunnelling conductance give
predictions for quasiparticle charge $e^*$ and the tunnelling particles
interaction parameter \cite{PhysRevB.90.075403}, $g$. There is still no obvious
match to the theoretical predictions of the various candidate states for the
edges in the $\nu=5/2$ system \cite{PhysRevB.88.085317} and even distinguishing
whether the state should display Abelian or non-Abelian statistics is not
obvious.

The RG approach by Kane and Fisher is based on a 1D lattice model and it is
assumed that the edge states of the FQHE will display similar behaviour so this
approach is frequently used as a base model for theoretical predictions on
transport properties of the FQHE. There are important differences between the
Luttinger liquid model used for the perturbative RG analysis by Kane and Fisher
and the FQH edge states. The electron field operators in the Luttinger liquid
model can be derived microscopically from the 1D Hamiltonian describing a system
of interacting electrons. It is not the same for the FQH edge states since in
this case, the edge states result from a two-dimensional system of
electrons in a strong magnetic field, thus the low-energy effective theory is
obtained by projecting the FQH states onto the space of low energy edge
states. The perturbative RG approach that works so well for the lattice model
Luttinger liquid cannot be extended straightforwardly to the FQH edge states
since it relies on the fact that interactions between electrons can be treated
perturbatively. Switching off the electron-electron interactions in the FQHE
destroys the effects existence. So how do these differences affect the
formulation of a chiral Luttinger liquid as compared to that of a conventional
Luttinger liquid?

It is already understood that the low energy projection of the edge states in
the FQHE do not display exactly the same behaviour as the Luttinger liquid. One
example is the low energy projection of the electron field operator. In the
Luttinger liquid the anti-commutation relation for two spatially separated
electron fields is given by a delta function, the same behaviour of similar
fields in the edges states of a FQH system is not observed. There is another
issue with the locality of the effective tunnelling Hamiltonian for a
quasiparticle being transferred between two disconnected edges in the system. In
the FQHE the tunnelling Hamiltonian takes a similar form to that of the
tunnelling Hamiltonian in the conventional Luttinger liquid. I.e. the operator
consists of creating a particle in one of the QH edge states and annihilating a
particle in the opposing edge \cite{PhysRevLett.71.4381,PhysRevLett.74.3005}. Without the perturbative RG analysis at our disposal
for FQH states there is no guarantee that the effective theory tunnelling
operators will be local. For the Luttinger liquid model however, local operators
in the microscopic theory are guaranteed to remain local in the effective theory
using the Kadanoff coarse graining procedure.

The problem of the locality of the tunnelling Hamiltonian has been investigated
and for a FQH system containing multiple quantum point contacts. It was observed
that the tunnelling operators at one of the QPC's did not commute with the
tunnelling operator at a different QPC, independent on the magnitude of their
spatial separation \cite{PhysRevB.65.153304}. To impose the expected locality
which the quasiparticle tunnelling operators should adhere, the effective
quasiparticle operators had additional Klein factors included in their
representation
\cite{PhysRevB.65.153304,PhysRevLett.90.226802,PhysRevLett.86.4628,PhysRevB.74.045319,PhysRevLett.91.196803,PhysRevB.59.15694,PhysRevB.70.195316}. The
addition of the Klein factors adds an extra phase to the quasiparticle fields
and it is reasoned that this is a statistical phase which is gained during a
tunnelling event between two disconnected edges. This statistical phase is a
result of the fractional statistics obeyed by the quasiparticles. Including
Klein factors results in the effective quasiparticle tunnelling operators at two
different QPC to commute with one another. Results on observables such as
tunnelling currents are greatly dependent on the inclusion of these Klein
factors (for example compare work by Law \textit{et al.}
\cite{PhysRevB.74.045319} with Jonckheere \textit{et al.}
\cite{PhysRevB.72.201305}).

In this work the low-energy projection of the tunnelling Hamiltonian is
investigated for the quantum Hall edge states by making comparisons to
microscopic calculations. In particular the form of the tunnelling Hamiltonian
is tested for a quasiparticle and an electron in the $\nu=1/3$ FQH state by
comparing it to exact microscopic calculations for the zero modes of the edge
states. From the perturbative RG analysis on the
conventional Luttinger liquid model, for a sufficiently weak barrier at
$\nu=1/3$, the only relevant tunnelling should be that of single Laughlin
quasiparticle with charge $e^* = \nu e$. Electron, and any combination of
multiple quasiparticle tunnelling should be suppressed in the weak back
scattering regime. Without any initial conditions to impose from a microscopic
theory it is not obvious at exactly what energy scales in the FQHE that these
higher-order tunnelling terms become irrelevant. From microscopic calculations
presented here however, it can be deduced at what (if at all) system sizes the
electron and quasiparticle tunnelling amplitudes become comparable. The
microscopic calculation for the tunnelling of particles is carried out
numerically using Monte Carlo where we focus on the simple case of investigating
how an impurity, when placed between two disconnected edges in the bulk, affects
the groundstate system. The effective theory tunnelling model already makes
strong predictions about the way these zero mode matrix elements should behave
for increasing system size and it is this relationship in particular that is
tested. We find that the microscopic computations are in good agreement with the
model so long as a specific type or ordering of operators is used in the
tunnelling Hamiltonian. If one chooses to use normal ordered operators, which of
course is convenient, then a normalisation factor dependent on the system size
should be included. Such a normalisation factor would also make the tunnelling
operator local.

The layout of this report is as follows; in section I. the geometry of the FQH
system is introduced and the specifics of the chiral Luttinger liquid formalism
is reviewed. The predictions of the sine-Gordon model for the specific geometry
considered in this work are also calculated. Section II. discusses the Monte
Carlo (MC) calculations whilst in the final section the MC results are presented
and comparisons are made to the effective low energy model of the tunnelling
operators.

\section{Overview of the Chiral Luttinger Liquid and the Boundary Sine-Gordon Model  Predictions}

In this work, only FQH states occupying the lowest Landau level are
considered. Since the exact ground state of a FQH system is not known, we
instead use the Laughlin wavefunction which has been shown to be a good
approximation for the system. This wavefunction is an exact solution for
an Hamiltonian in terms of Haldane's pseudo-potentials
\cite{PhysRevLett.51.605}. For a droplet of $N$ electrons in the lowest Landau
level with filling factor $\nu = m^{-1}$, where $m$ is an odd integer, Laughlin's
wavefunction \cite{laughlinPA} has the form


\begin{equation} \label{laughlinstate}
\Psi_{N} = \displaystyle \prod_{k=1}^N{ \left( e^{-\frac{|z_k|^2}{4l}}\right)}
\prod_{i<j}^N{(z_i-z_j)^m} 
\end{equation}

where $l$ is the magnetic length which, for the remainder of this work will be
set to unity. The $\{z_i\}$ correspond to electron positions and the FQH droplet
has a radius $R=\sqrt{2 m N}$. It is noted that this wavefunction is not
normalised. To create a second edge in our system such that tunnelling across
the bulk can be observed, a macroscopic hole is inserted in the centre of the
droplet. This macroscopic hole is created by inserting $M$ quasiholes at
$z=0$. Mathematically, this corresponds to an extra product over particle
coordinates in the wave function.

\begin{equation} \label{laughlin_mysystem}
\Psi_N^M = \prod_{k=1}^N{ \left( e^{-\frac{|z_k|^2}{4l}} z_k^M \right)}
\prod_{i<j}^N{(z_i-z_j)^m} \equiv \left| N,M \right\rangle
\end{equation}

\begin{figure}[htb]
\begin{center}                        
\includegraphics[scale=0.7]{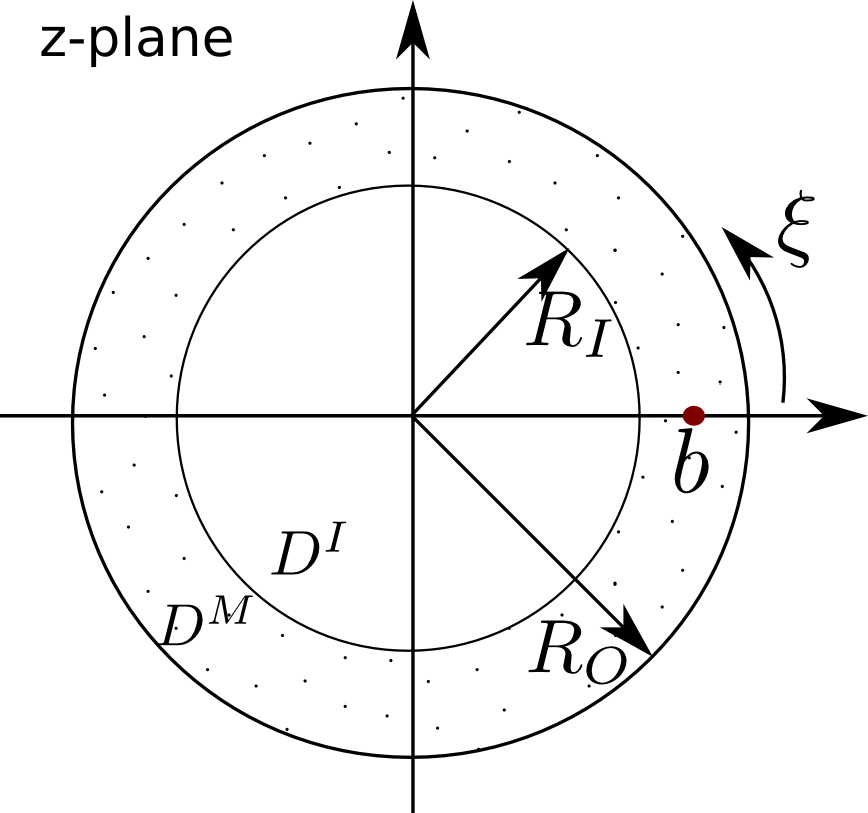}
\caption{Schematic diagram for the FQH device. Charges are confined to the
  domain $D^M$ on the complex plane. The radii of the inner and outer
  edge of $D^M$ are given by $R_I$ and $R_O$. The width of domain $D^M$ is such
  that $R_O - R_I \sim 4$. Domain $D^I$ corresponds to the area of the macroscopic
  hole created by inserting Laughlin quasiholes at $z=0$. An impurity is placed
  at position $b$ along the positive real axis.}
\label{zero_mode_system}
\end{center}
\end{figure}

The bra-ket notation has been introduced to simplify expressions in future
calculations. A schematic diagram of the device is shown in Figure
\ref{zero_mode_system}. Now the electrons are confined to a domain $D^M$ on the
complex plane. The hole created at $z=0$ has a radius $R_I = \sqrt{2M}$ and the
outer radius is given by $R_O = \sqrt{2mN+2M}$. At the inferaces of $D^M$ there
is a sharp decrease of particle density to zero in the large $N$ limit.
\newline

To encourage tunnelling between the edges of the system, an impurity is placed
inside the bulk at position $b$. The potential $\hat{V}$ of this impurity has
the form

\begin{equation} \label{imp_pot}
\hat{V} = u \sum_{i=1}^N \delta^{(2)}(z_i - b).
\end{equation}

The parameter $u$ corresponds to the strength of the potential which will be set
to unity and we assume that $|b| = (R_O+R_I)/2$. If the ring of the bulk is
sufficiently thick then adding the impurity to the system will have no effect on
the edges due to a finite correlation length in the bulk on the order of the
magnetic length. Therefore it is assumed that for all $N$ the width
of the system ($R_O-R_I$) is constant and narrow enough such that both edges are
affected by the impurity.

As implied in the introduction, it is the edges of the device, consisting of two
counter propagating chiral Luttinger liquids that are of most interest for this
work. A successful method of the describing the effective low-energy physics of
a Luttinger liquid uses the method of bosonization. There are various approaches
to deriving the electron fields and in this report a similar notation is used to
the formalism in \cite{vadim1} and \cite{LFS_ll_annulus}. Here, operators are
projected onto the subspace of edge excitations which defines the zero mode
operators and Boson creation and annihilation operators for quasiparticles. It
is assumed that the edges are sufficiently far apart that they do not interact
away from the impurity and thus each edge has its own electron field operators.

In the effective theory of low-energy excitations the transverse positions to
the edges of the FQH device are unimportant since we assume that the domain
$D^M$ is a very narrow ring in comparison to the size of its outer radius $R_O$,
therefore our particle fields depend only on the longitudinal coordinate $\xi$,
where $\xi \in [0, 2 \pi R]$ and $R$ is the radius at which the transfer of
charges between the edges takes place, i.e. $R \equiv |b|$. One can write the
bosonization formulae for general field operators $\psi_p$ in the low-energy
effective theory, where $p$ is some integer and $\psi_p$ corresponds to the
annihilation operator of particle with charge $e^* = (p/m)e$ in the Laughlin
state $\nu=(1/m)$. Therefore $\psi_{p=1}$ corresponds to the field for a single
Laughlin quasiparticle with charge $e^*=(1/m)e$ and $\psi_{p=m}$
corresponds to $m$ quasiparticles, or equivalently a single electron field.

\begin{eqnarray} \label{p_field_op_longitud}
\psi^\dagger_{p,O}(\xi) &=& e^{i \frac{p}{m}\varphi_N}
e^{p\phi_O(\xi)} \nonumber \\
\psi_{p,O}(\xi) &=& e^{-p\phi_O(\xi)}
e^{-i\frac{p}{m}\varphi_N} \nonumber \\
\psi^\dagger_{p,I}(\xi) &=& e^{i\frac{p}{m}\varphi_N} e^{ip\varphi_M}
e^{p\phi_I (\xi)} \nonumber \\
\psi_{p,I}(\xi) &=& e^{-p\phi_I (\xi)}
e^{-i\frac{p}{m}\varphi_N} e^{-ip\varphi_M} 
\end{eqnarray}

The subscript "I" corresponds to an inner boundary operator and "O" to an outer
boundary operator. The fields $\phi_{O/I}$ are bosonic and are given by

\begin{eqnarray} \label{el_bose_field}
\phi_O(\xi) &=& -\frac{i\xi}{R}(\theta_N +\frac{\theta_M}{m}) +\sum_{k>0} \sqrt{\frac{1}{mk}} \left(
  e^{-ik\frac{\xi}{R}} a_k^\dagger - e^{ik\frac{\xi}{R}} a_k \right) \nonumber \\
&=& \phi_O^0 + \phi_O^+ - \phi_O^- \nonumber \\
\phi_I(\xi) &=& -\frac{i\xi}{mR}\theta_M +\sum_{k>0} \sqrt{\frac{1}{mk}} \left(
  e^{ik\frac{\xi}{R}} a_{-k}^\dagger - e^{-ik\frac{\xi}{R}} a_{-k} \right) \nonumber \\
&=& \phi_I^0 + \phi_I^+ - \phi_I^-. 
\end{eqnarray}

The short hand notation of decomposing the the field $\phi$ into zero mode parts
($\phi^0$), creation operator parts ($\phi^+$) and annihilation operator parts
($\phi^-$) will be used in subsequent calculations. The $a^\dagger_{\pm
  k}$ and $a_{\pm k}$ are edge excitation creation and annihilation operators for a
given excitation orbital $k$ which may correspond to either the outer ("+") or
inner ("-") boundary. Operators $\theta_{N/M}$ and $e^{i \varphi_{N/M}}$ are
conjugate zero mode operators and act as follows;

\begin{eqnarray}
e^{i \varphi_N} \left| N, M \right\rangle &=& \left| N+1, M
\right\rangle \nonumber \\
\theta_N \left| N, M \right\rangle &=& N \left| N, M \right\rangle \nonumber \\
e^{i \varphi_M} \left| N, M \right\rangle &=& \left| N, M+1
\right\rangle \nonumber \\
\theta_M \left| N, M \right\rangle &=& M \left| N, M \right\rangle \nonumber 
\end{eqnarray}

where $\left| N, M \right\rangle$ is the Laughlin groundstate for a system of
$N$ electrons and $M$ quasiholes. In this effective theory the states $\left| N, M
\right\rangle$ are normalised and orthogonal such that $\left\langle N, M| N',
  M' \right\rangle = \delta_{N,N'} \delta_{M,M'}$ in the large $N$ limit
\cite{laughlinPA}. The operators for the outer and inner
boundary are independent and thus all commute with one another. In summary, a
list of useful commutation relations are;

\begin{eqnarray} \label{useful_commutators}
\left[\theta_x, e^{i \varphi_y} \right] &=& e^{i \varphi_x} \delta_{x,y}
  \quad \textrm{where} \quad x \, \, \textrm{or} \, \, y = N \, \, \textrm{or}
  \, \, M \nonumber \\
\left[a_{\pm p}, a^\dagger_{\pm k} \right] &=& \delta_{p,k}  \nonumber \\
\left[a_{\pm p}, a^\dagger_{\mp k} \right] &=& 0
\end{eqnarray} 

With the low-energy, effective theory field operators defined, one can now
discuss the effect of the impurity placed in the bulk. The form of the
tunnelling operator used in all the referenced work in the introduction takes
the form

\begin{equation} \label{general_TH}
H_T = \sum_{p=1}^{\infty} A_p(\xi)
\end{equation}

where $\xi$ is the longitudinal position at which tunnelling occurs along the
boundaries and operators $A_p$ transfer a number of $p$ quasiparticles from the
inner to the outer boundary. From the form of $H_T$ one can see it is possible
for the tunnelling of any number of quasiparticles across the bulk. For $p=1$,
$A_P$ describes the tunnelling process for a single quasiparticle and for $p=m$
the tunnelling process is for that of an electron in FQH state $\nu=1/m$. For
Laughlin states however, it can be seen from a microscopic calculation (see
section II. for details) that the transfer of $p>m$ quasiparticles is identically
zero. Whether this behaviour should be extended to the exact wavefunctions of
quantum Hall states is doubtful and thus experiments on FQH systems that measure
a $p>m$ tunnelling process would be extremely interesting and a good measure for
the preciseness of Laughlin's wavefunction as a description FQHE.

Here two types of tunnelling processes across the bulk are considered, the first being a single
quasiparticle tunnelling and the second will be electron tunnelling or
equivalently $m$ quasiparticles tunnelling. These processes are described by
$A_{p=1}$ and $A_{p=m}$ respectively. In general operators $A_p$ have the form;

\begin{equation} \label{tunnelling_op_el}
A_p (\xi) = t_p \left( \psi_{p,O}^\dagger(\xi) \psi_{p,I}(\xi) + \textrm{h.c.}
\right).
\end{equation}

where $\xi$ is the longitudinal position at which the tunnelling occurs on the
boundaries and $t_p$ is a parameter that cannot be calculated analytically given
the microscopic theory. In this work the scaling behaviour of the parameters
$t_p$ will be investigated as system size $N$ is varied. This can be achieved by
looking at the zero mode matrix elements of the tunnelling elements.

\begin{eqnarray} \label{zm_tunnelling_me_p}
\left\langle A_p (\xi) \right\rangle &=& \left\langle N, M \right| A_p(\xi)
\left| N', M' \right\rangle \nonumber \\
 &=& t_p\left\langle
  N, M \right| \psi_{p,O}^\dagger(\xi) \psi_{p,I}(\xi) \left| N', M'
\right\rangle + \textrm{h.c.} \nonumber \\
&=& t_p\left\langle
  N, M \right| e^{i \frac{p}{m}\varphi_N} e^{p \phi_O(\xi)} e^{-p\phi_I(\xi)} e^{-i
  \frac{p}{m}\varphi_N} e^{-ip\varphi_M} \left| N', M' \right\rangle + \textrm{h.c.}
\end{eqnarray}

The first term in (\ref{zm_tunnelling_me_p}) describes a process where $p$
particles move from the inner boundary to the outer boundary and for the second
term this process is reversed. Only one term is needed here so the Hermitian
conjugate term can be neglected.  Next the ordering of the operators is considered
which will in turn affect the behaviour of the parameter $t_p$. Here, two types
of ordering are considered with the first being the usual definition of normal
ordering defined by

\begin{equation} \label{norm_ordering}
e^{\phi} \rightarrow :e^{\phi}:\, = e^{\phi^0}
e^{\phi^+} e^{-\phi^-}
\end{equation}

The matrix element calculated using this ordering is denoted by
$\left\langle :A_p(\xi): \right\rangle$ and is straightforward to calculate. 

\begin{equation} \label{zm_tunnelling_me_elNO} \left| \left\langle :A_p(\xi):
    \right\rangle \right| = \left| \left\langle N, M \left| e^{i
          \frac{p}{m}\varphi_N} :e^{p \phi_O(\xi)}: :e^{-p\phi_I(\xi)}: e^{-i
          \frac{p}{m}\varphi_N} e^{-ip\varphi_M} \right| N', M' \right\rangle
  \right| = |t_p| \delta_{N,N'} \delta_{M,M'-p}
\end{equation}

Note that the absolute values of matrix elements are taken to get rid of
unnecessary phase terms. Therefore if normal ordering is used, only $t_p$
contributes to system size dependence. On the other hand, a different result is
obtained by choosing the following ordering for the operators,

\begin{equation} \label{op_ordering}
e^{\phi} \, = e^{\phi^0}
e^{\phi^+} e^{-\phi^-} e^{-\frac{1}{2}[\phi^0,
  \phi^+-\phi^-]}
e^{-\frac{1}{2}[\phi^+,-\phi^-]} = e^{\phi^0}
e^{\phi^+} e^{-\phi^-} e^{-\frac{1}{2}\sum_k(\frac{m}{k})}
\end{equation}

where the following commutation relations have been used; $[\phi^0_x(\xi),
\phi^+_x(\xi)-\phi^-_x(\xi)] = 0$ and $[\phi^+_x(\xi),\phi^-_x(\xi)] =- \sum_k
(m/k)$ where $x = I$ or, $x=O$. These relations are straightforward to calculate using the correlators
listed in (\ref{useful_commutators}). Proceeding with the matrix element
(\ref{zm_tunnelling_me_p});

\begin{eqnarray} \label{zm_tunnelling_me_el2} \left| \left\langle A_p(\xi)
    \right\rangle \right| &=& |t_p| \exp\left(-\sum_{k>0}\frac{p^2}{mk} \right)
  \left| \left\langle N-1, M \right| e^{p \phi_O^0} e^{p\phi_O^+} e^{-p\phi_O^-}
    e^{-p\phi_I^0} e^{-p\phi_I^+} e^{p\phi_I^-} \left| N'-1, M'-m \right\rangle
  \right|
  \nonumber \\
  &=& |t_p| \exp\left( -\sum_{k>0}\frac{p^2}{m k} \right) \delta_{N,N'}
  \delta_{M,M'-p} 
\end{eqnarray}

The final step is to carry out the sum over $k$ in
(\ref{zm_tunnelling_me_el2}). There are two natural cutoffs that can be
considered for the low-energy limit of the system. The first is related to the
breakdown of the Boson creation and annihilation operators in the Luttinger
liquid theory which are only independent operators for $k \leq N$. Therefore $N$
could be a valid cutoff for the sum. However there is also an obvious limit on
the energy of the edge excitations which is related to the bulk energy gap. If
quasiparticles have an energy larger than the bulk gap energy $\Delta$ then they
are able to travel through the bulk destroying the physics of the Luttinger
liquid edge states in the FQHE. The dispersion relation for the Bose
excitations is linear at the edge and so the maximum momentum for a
quasiparticle is $p = \Delta / v$ where $v$ is the quasiparticle velocity. The
momentum corresponding to a given edge orbital $k$ is also given by $p = \hbar k
/R$ and therefore the maximum value of the orbital according to the bulk energy
gap is $k \leq R \Delta /(v \hbar) \equiv \Lambda$. Recall that $R \equiv |b|$,
the point at which tunnelling occurs, and taking into account the constraint on
$M$ that for any $N$ one must always have $R_O-R_I \sim 4$, then the magnitude
of the impurity, $b$ is on the order of $N$ in the large $N$ limit. Thus the two
cutoffs are essentially equivalent. In this work we use $\Lambda \sim R \equiv
|b|$ as the cutoff, which has more of an apparent, physical meaning.

The sum over $k$ in (\ref{zm_tunnelling_me_el2}) is performed using $\Lambda
\sim R \equiv |b|$ as a soft cutoff such that $\sum_{k>0} k^{-1} \rightarrow
\sum_{k>0} k^{-1} e^{-k/R} = - \ln(1-e^{-1/R})$. To get an approximate idea of
the behaviour of this sum in the large $N$, or equivalently large $R$ limit, the
exponential in the logarithm can be expanded to give $-\ln(1-e^{-1/R}) \sim \ln
R$. Thus for a log-log plot of the amplitude of (\ref{zm_tunnelling_me_el2})
versus particle number, one would expect the gradient to be $\sim -p^2/m$ for
large enough $R \equiv |b|$. For now we will keep the exact form of the sum so
that the final expression of the tunnelling matrix elements is,

\begin{equation} \label{zm_tunnelling_me_el3} \left| \left\langle A_p(\xi)
    \right\rangle \right| =|t_p| \left(1-e^{-\frac{1}{R}}
  \right)^{\frac{p^2}{m}} \delta_{N,N'} \delta_{M,M'-p}
\end{equation}

From (\ref{zm_tunnelling_me_el3}); the zero mode tunnelling matrix elements have
a dependence on the system size $N$, which for non-normal ordering given by
(\ref{op_ordering}) is not from the parameter $t_p$. If $A_p$ for ordering
(\ref{op_ordering}) is local then we can assume all the system size dependence
originates from the sum $\sum_k k^{-1}$ and $t_p$ is constant for all
$N$. Ideally the tunnelling operators should remain local since the impurity
placed between the edges in the FQH device should only affect carriers in its
vicinity and not the remainder of the system. The correlator of a local operator
$\left\langle A_p(\xi) A_p (\xi') \right\rangle$ should be independent of system
size for $\xi$ and $\xi'$ sufficiently far apart, also $\xi$, $\xi'$ should be
sufficiently far away from the edges of the system. 

\begin{eqnarray} \label{qp_correlator}
 \left\langle A_p(\xi) A_p(\xi') \right\rangle &=& 
  \left\langle N,M \right| A_p(\xi) A_p(\xi') \left| N', M' \right\rangle \nonumber \\
&=& 2 \exp \left\{ \frac{2 p^2}{m} \sum_{k>0} k^{-1} \left( \cos \left[ \frac{k}{R}(\xi'
    -\xi)\right]-1 \right) \right\}\cos \left[\frac{p}{R} \left(N -\frac{p}{m}
\right)(\xi'-\xi) \right] 
\end{eqnarray}

Only matrix elements that are kept in the above expression are those with
$M'=M$ and $N'=N$ which satisfies the definition of the tunnelling operator
$A_p$. 

To finish the calculation, a soft cutoff is used ($\Lambda \equiv R$) to
calculate the sum in (\ref{qp_correlator}) which is the same approach as was
used to calculate similar sums earlier for the tunnelling matrix elements. Using
this cutoff and taking the limit $|\xi-\xi'|/R << 1$ gives the final result for
the correlator calculation which has, as required, no system size independence.

\begin{equation} \label{qp_correlator2} 
\left\langle A_p(\xi) A_p(\xi') \right\rangle = \frac{2}{ \left(1+(\xi' -\xi)^2
  \right)^{\frac{p^2}{m}}} \cos \left[\frac{p}{R} \left(N -\frac{p}{m} 
\right)(\xi'-\xi) \right]
\end{equation}

The correlator of the tunnelling operator for normal ordering, $\left\langle
  :A_p(\xi): :A_p(\xi'): \right\rangle$ does have system size dependence and
therefore tunnelling operators $A_p$ defined by ordering (\ref{op_ordering}) are
preferred.  To summarise this section; expressions for the zero mode matrix
tunnelling elements have been calculated for both normal ordered operators
(\ref{zm_tunnelling_me_elNO}) and for non-normal ordered operators
(\ref{zm_tunnelling_me_el3}). In the normal ordered case there is system size
dependence which must be encoded in the parameter $t_p$. For the non-normal
ordered tunnelling operators, this size dependence comes from the calculation of
the matrix elements and the operator algebra itself. Since the non-normal
ordered tunnelling matrix elements show explicit size dependence and were
obtained from local tunnelling operators, we concentrate
on these when making comparisons with the microscopic theory discussed in the
next section.


\section{Microscopic Calculations for Tunnelling Matrix Elements}

The results for the effective theory tunnelling operators calculated in the
previous section should also manifest from a microscopic theory. Here it is
assumed that the Laughlin state is exact wavefunction for the FQH
system with filling factor $\nu=1/m$, where $m$ is odd. The microscopic
expression for the zero mode tunnelling matrix elements are given by

\begin{eqnarray} \label{tme_qpChi} 
\left\langle V \right\rangle_\chi &\equiv& \frac{\left\langle N,M 
    \left| V \right|  N, M+\chi \right\rangle} {\sqrt{
        \left\langle N, M | N, M \right\rangle
       \left\langle N, M+\chi | N, M+\chi \right\rangle }} \nonumber \\
&=&
\frac{ \displaystyle \int \prod_{k=1}^{N}
 \left( \textrm{d}^2z_k \, e^{-\frac{|z_k|^2}{2}}|z_k|^{2M}z_k^\chi
 \right) \prod_{i<j}^{N}|z_i-z_j|^{2m} \sum_{n=1}^N \delta^{(2)} (z_n-b) }
{ \displaystyle \left[ \int \prod_{k=1}^{N}
 \left( \textrm{d}^2z_k \, e^{-\frac{|z_k|^2}{2}}|z_k|^{2M}
 \right) \prod_{i<j}^{N}|z_i-z_j|^{2m} \cdot \int \prod_{k=1}^{N}
 \left( \textrm{d}^2z_k \, e^{-\frac{|z_k|^2}{2}}|z_k|^{2(M+\chi)}
 \right) \prod_{i<j}^{N}|z_i-z_j|^{2m} \right]^{\frac{1}{2}}}
\end{eqnarray}

This matrix element describes a process in which a number of $\chi$
quasiparticles is transferred from the inner boundary to the outer boundary due
to the impurity potential $\hat{V}$ given in (\ref{imp_pot}). Therefore $\chi =
1$ corresponds to quasiparticle tunnelling and $\chi = m$ to electron tunnelling
for some Laughlin state $\nu = 1/m$. Thus $\left\langle V \right\rangle_\chi$
and $\left\langle A_\chi \right\rangle$ in (\ref{zm_tunnelling_me_el3})
physically describe the same process and so if the effective theory is a good
description for the FQH edge states then it is expect $\left\langle A_\chi
\right\rangle \equiv \left\langle V \right\rangle_\chi$. Note that the
denominator in (\ref{tme_qpChi}) is needed to correctly normalise the elements
since the states are now the Laughlin wavefunctions given by
(\ref{laughlin_mysystem}). To shorten notation inside the integrals,
$|\Psi_N^M|^2$ will be used to denote the absolute value of Laughlin's
wavefunction squared and the integration variables are shortened to; $\int_{N}
\equiv \int \prod_{k=1}^N \mathrm{d}^2 z_k$. With this notation

\[
\left\langle V \right\rangle_\chi = \frac{\displaystyle
   \int_{N-1} |\Psi_{N}^M|^2 \prod_{i=1}^{N-1}
  z_i^\chi \sum_{n=1}^N \delta^{(2)}(z_n-b)}{ \displaystyle \sqrt{\int_N
    |\Psi_{N}^M|^2 \cdot \int_N |\Psi_{N}^{M+\chi}|^2}}
\]

The delta-function in the numerator of (\ref{tme_qpChi})
allows one of the variables in the integral to be integrated out. Using the fact
that the integrals appearing in each term of the sum are symmetric with respect
to the exchange of integration variables, $\left\langle V \right\rangle_\chi$ can
be written as 

\begin{equation} \label{tme_qpChi2} \left\langle V \right\rangle_\chi =
  \frac{\displaystyle N e^{-\frac{|b|^2}{2} |b|^{2M}} b^\chi \int_{N-1}
    |\Psi_{N-1}^M|^2 \prod_{i=1}^{N-1} z_i^\chi |z_i-b|^{2m}}{ \displaystyle
    \sqrt{\int_N |\Psi_{N}^M|^2 \cdot \int_N |\Psi_{N}^{M+\chi}|^2}}
\end{equation}

As far as this author is aware, there are no known analytic solutions
to (\ref{tme_qpChi}) and so the computations are performed numerically. The
overlap integrals in (\ref{tme_qpChi}) for the free fermion case can be
calculated analytically and thus provide a good check for the numerical methods
used in this work. An effective numerical method to calculate these overlap
integrals is by using Monte Carlo (MC) simulations. As already
discussed, this method seems quite natural since the norm of the wavefunction
can be considered as a partition function for a 2D Coulomb plasma
\cite{laughlinPA} allowing statistical averages of operators to be calculated
with a probability distribution analogous of the Boltzmann distribution of the
plasma. All the MC simulations in the present work were carried out using the
Metropolis algorithm \cite{metropolisAlg}.

To directly use the MC method on the integral in the numerator of
(\ref{tme_qpChi}) is difficult due to there being the product over all particles
of the form $z_i^\chi$. This product introduces a phase problem to the
calculation since the MC measurements will fluctuate
between positive and negative values and the convergence of the simulation will
be slow. Two successful methods to overcome this phase problem have been
found. The effectiveness of these methods strongly depends on the value of
$\chi$ and the first of the methods to be discussed is appropriate for small
values of $\chi$ whereas the second method can only be used $\chi = m$,
i.e. for the case of an electron tunnelling across the bulk.

\subsection{Phase Problem Solution: Method 1 for  $\chi \leq 1$}

The first method of overcoming the phase problem in the integral
(\ref{tme_qpChi}) is by using the cumulant expansion. It will be seen later that
only for a small value of $\chi$ can the particular cumulant expansion of
interest be calculated reliably. First the integral in the numerator should be
expressed in a more convenient way. To do this it is noted that part of the
numerator of (\ref{tme_qpChi2}) can be re-written as;

\[ 
\prod_{i=1}^{N-1} |z_i-b|^{2m} z_i^\chi = \left( \frac{-b}{|b|}
\right)^{\chi(N-1)} \prod_{i=1}^{N-1} |z_i - b|^{2m} |z_i|^\chi
\Theta_\chi(z_i,b) 
\]

where

\begin{equation} \label{theta_func}
\Theta_\chi(z_i,b) = \left[ \frac{\left(1-\frac{z_i}{b} \right)
    \left(1-\frac{\bar{b}}{\bar{z}_i} \right)}{\left|1-\frac{z_i}{b} \right|
    \left|1-\frac{\bar{b}}{\bar{z}_i} \right|}\right]^\chi.
\end{equation}

The advantage of the function $\Theta_\chi(z,b)$ is that its cumulant expansion
can be calculated with respect to some probability distribution using a MC
simulation with a relatively quick convergence. Indeed if we choose the
probability distribution such that

\begin{equation} \label{theta_av} \left\langle \prod_{i=1}^{N-1}
    \Theta_\chi(z_i,b) \right\rangle_\varphi = \frac{\displaystyle \int_{N-1}
    |\Psi_{N-1}^M|^2 \prod_{i=1}^{N-1} |z_i|^\chi \, |z_i-b|^{2m}
    \Theta_\chi(z_i,b)} { \displaystyle \int_{N-1} |\Psi_{N-1}^M|^2
    \prod_{i=1}^{N-1} |z_i|^\chi \, |z_i-b|^{2m}}
\end{equation} 

then the average over $\Theta(z,b)$ in (\ref{theta_av}) can be related the
tunnelling matrix elements in (\ref{tme_qpChi}) via

\begin{equation} \label{V_chi1} \left\langle V \right\rangle_\chi =
  \left(\frac{-|b|}{b} \right)^{\chi(N-1)} \left\langle \prod_{i=1}^{N-1}
    \Theta_\chi(z_i,b) \right\rangle_\varphi N e^{\frac{|b|^2}{2}} |b|^{2M}
  b^\chi \sqrt{h_1 h_2}
\end{equation} 

where

\begin{equation} \label{h_1_def}
h_1 = \frac{ \displaystyle
    \int_{N-1} |\Psi_{N-1}^M|^2 \prod_{i=1}^{N-1} |z_i|^\chi |z_i-b|^{2m}} { \displaystyle
    \int_N |\Psi_N^M|} 
\end{equation}
\begin{equation} \label{h_2_def}
h_2 = \frac{ \displaystyle
    \int_{N-1} |\Psi_{N-1}^M|^2 \prod_{i=1}^{N-1} |z_i|^\chi |z_i-b|^{2m}} { \displaystyle
    \int_N |\Psi_N^{M+\chi}|}. 
\end{equation}

Integrals $h_1$ and $h_2$ are relatively trivial MC integrals to calculate. As
already mentioned it is the cumulant expansion of $\Theta(z,b)$ in
(\ref{theta_av}) that gets rid of the phase problem. In order to see why this is
so consider $\Theta(z,b)$ written in terms of an exponential.

\begin{equation} \label{ln_theta} e^{\ln \Theta(z,b)} = \exp \left( i \chi
    \,\textrm{Arg}\left[\left(1-\frac{z}{b}
      \right)\left(1-\frac{\bar{b}}{\bar{z}} \right) \right] \right)
\end{equation}

The first couple of terms for the cumulant expansion of some function
$\left\langle e^G \right\rangle$ is 

\begin{equation} \label{cumu_exp}
\left\langle e^G \right\rangle = e^{\left\langle G \right\rangle +
  \frac{1}{2!} \left( \left\langle G^2 \right\rangle - \left\langle G
  \right\rangle^2 \right)+ \frac{1}{3!}\left(2 \left\langle G \right\rangle^3 -
  3\left\langle G \right\rangle \left\langle G^2 \right\rangle + \left\langle G^3
  \right\rangle \right)+\cdots }
\end{equation}

After substituting (\ref{ln_theta}) into the expansion of (\ref{cumu_exp}) we
notice the following; the mean field of $G$ is essentially an average over an
angle, which is zero. Secondly, sequential terms in the expansion increases by a
factor of $\chi$. For the cumulant expansion to be used reliably, the higher
order terms must quickly decay to zero. For $\chi > 1$ then this is not the
case. However for $\chi = 1$ the cumulant expansion of $\left\langle
  \prod\Theta(z_i,b) \right\rangle$ is well behaved and can be used to a good
accuracy to calculate the matrix elements in (\ref{tme_qpChi}).

\subsection{Phase Problem Solution: Method 2 for $\chi=m$}

The second method of avoiding the phase problem in (\ref{tme_qpChi}) relies on
the fact that for the special case $\chi = m$, then (\ref{tme_qpChi}) can be
written in terms of real valued functions. The numerator of (\ref{tme_qpChi})
(sticking to general $\chi$ for the moment) can be written as follows,

\begin{eqnarray} \label{N3qp_tme_chi} \left\langle N,M \left|
  V \right| N,M+\chi \right\rangle &=& N e^{-\frac{b^2}{2}}b^{2M+\chi}
  \int \prod_{i=1}^{N-1}{\left( \mathrm{d}^2 z_i e^{-\frac{|z_i|^2}{2}} |z_i|^{2M} z_i^\chi
      |z_i - b|^{2m}\right)} \prod_{i<j}^{N-1} |z_i-z_j|^{2m} \nonumber
  \\
  &=& N e^{-b^2/2}b^{2M+\chi} \int_{N-1} \left|\Psi_{N-1}^M \right|^2
  \prod_{i=1}^{N-1}{ |z_i|^{2\chi} \bar{z}_i^{m-\chi} \left(z_i-b \right)^m
      \left(1- \frac{\bar{b}}{\bar{z}_i} \right)^m}.
\end{eqnarray}

By writing the numerator of $\left\langle V \right\rangle_\chi$ in this form, we
see that under a global rotation this integral will transform non-trivially for
$m-\chi < 0$. Therefore for $\chi >m$ the tunnelling matrix elements $\left\langle
V \right\rangle_\chi$ are zero for Laughlin wavefunctions. Under a similar
global rotation argument when $\chi = m$, the numerator of $\left\langle V
\right\rangle_m$ takes the particularly simple form

\begin{equation} \label{N3qp_tme} \left\langle N,M \left| V \right| N,M+m
  \right\rangle = N e^{-b^2/2}b^{2M+m} \left( -b\right)^{m(N-1)} \int_{N-1}
  \left|\Psi_{N-1}^{M+m} \right|^2.
\end{equation}

Therefore one can write the matrix element for the electron tunnelling as


\begin{equation} \label{3qp_tme_g1g2} \left\langle V \right\rangle_m = N
  e^{-\frac{b^2}{2}} (-b)^{m(N-1)} b^{2M+m} \sqrt{\frac{\Xi^2}{\tau}}.
\end{equation}

The integrals have all been symbolised by $\Xi$ and $\tau$ to shorten
notation and they have the form

\begin{eqnarray} \label{expression_Xi}
\Xi &=& \frac{ \displaystyle
    \int_{N-1} \left|\Psi_{N-1}^{M+m} \right|^2} { \displaystyle
    \int_N \left|\Psi_{N}^{M+m} \right|^2} \\
\tau &=& \frac{ \displaystyle
    \int_{N} \left|\Psi_{N}^{M} \right|^2} { \displaystyle
    \int_N \left|\Psi_{N}^{M+m} \right|^2} \nonumber 
\end{eqnarray}

Functions $\Xi$ and $\tau$ are simply ratios of the overlap of the groundstate
functions (\ref{laughlin_mysystem}) for various $N$ and $M$ values. The function
$\tau$ is a relatively trivial MC calculation, $\Xi$ is not so trivial
due to the differing number of integration variables in the numerator and the
denominator. To form a method to compute $\Xi$, consider the following average

\begin{equation} \label{F_average} \left\langle F \right\rangle_{N, M+m} =
  \frac{ \displaystyle \int_N \left| \Psi^{M+m}_{N}\right|^2
    \sum_{i=1}^{N}f(z_i)}{ \displaystyle \int_N \left|
      \Psi_{N}^{M+m}\right|^2 } 
  = N \frac{ \displaystyle \int_N \left| \Psi^{M+m}_{N}\right|^2
    f(z_N)}{ \displaystyle \int_N \left|
      \Psi^{M+m}_{N}\right|^2 } 
\end{equation}

The explicit form of $f(z)$ is chosen to be

\begin{equation} \label{count_f}
f(z_i) = \Theta(|z_i| - (R_O+d)) = \left\{
\begin{array}{ll}
1 & \quad \mathrm{for} \quad |z_i| \geq R_O+d \\
0 & \quad \mathrm{otherwise}
\end{array} \right.
\end{equation}

where $d$ is some distance added to the outer radius to be defined later. By
multiplying and dividing the average $\left\langle F \right\rangle_{N, M+m}$
by $\int_{N-1}|\Psi^{M+m}_{N-1}|^2$ then (\ref{F_average}) can be manipulated in
such a way as to contain the function $\Xi$. I.e.

\begin{equation} \label{f_average2} \left\langle F \right\rangle_{N, M+m}
  = N \cdot \Xi \cdot \frac{ \displaystyle \int_N \left| \Psi^{M+m}_{N}\right|^2
    f(z_N)}{ \displaystyle \int_{N-1} \left|
      \Psi^{M+m}_{N-1}\right|^2 } = N \, \Xi \, I_N
\end{equation}

The remaining integrals in the expression (\ref{f_average2}) have been labelled by $I_N$.

\begin{eqnarray} \label{IN_definition} 
I_N &=& \frac{ \displaystyle
    \int{\mathrm{d}^2z_N e^{-|z_N|^2/2} |z_N|^{2(M+m)} f(z_N)} \int_{N-1} \left|
      \Psi^\beta_{N-1}\right|^2 \prod_{i=1}^{N-1}|z_i - z_N|^{2m}}{
    \displaystyle \int_{N-1} \left| \Psi^\beta_{N-1}\right|^2 } \nonumber \\
  &=& 2\pi \int_{R_O+d}^\infty{\mathrm{d}r_N \,e^{-r_N^2/2} \,r_N^{2(M+m)
      +2m(N-1)}} \left\langle \prod_{i=1}^{N-1} \left| 1 - \frac{z_i}{z_N}
    \right|^{2m}
  \right\rangle_{N-1,M+m} \nonumber \\
  &=& 2\pi \int_{R_O+d}^\infty{\mathrm{d}r_N \, e^{-r_N^2/2} \, r_N^{2(M+m)
      +2m(N-1)}} \gamma(z_N)
\end{eqnarray}

The average over $N-1$ particles is labelled as $\gamma$, which is a function of
the $N$'th particle coordinate. Since it will be real valued the angle
integration over the $N$'th coordinate has been performed. The function
$e^{-r_N/2}r_N^{2(M+m)+2m(N-1)}$ is a rapidly decaying function away from the
outer boundary and thus if the lower integration limit of $r_N$ is sufficiently
larger than $R_O$, then one can use the following asymptotic approximation
$\gamma(z_N) \sim \gamma(R_O+d)$. This substitution makes the function
$\gamma(R_O+d)$ trivial to calculate using a MC simulation. Once $\gamma(R_O+d)$
is known, the integral $I_N$ is trivial to calculate.

Alongside the $\gamma$-function, the average $\left\langle F \right\rangle$ must
also be calculated using an MC simulation, which is also dependent on $d$, the
distance from the outer boundary. Since $\left\langle F \right\rangle$ simply
counts the number of particles beyond the point $R_O+d$ there is a trade-off as
to how large $d$ should be. The larger the value of $d$ the fewer particles
there will be to count since the particle density decreases sharply from the
outer boundary, but also for larger values of $d$, the more accurate the
asymptotic approximation, $\gamma(z_N) \sim \gamma(R_O+d)$, thus a trade off
must be made. With

\[
\Xi = \frac{\left\langle F \right\rangle_{N,M+m}}{N I_N}
\]

then the zero mode tunnelling matrix elements can be expressed in terms of the
MC integrals, for $\chi = m$ as follows,

\begin{equation} \label{NtoA_qp3}
\left\langle V \right\rangle_m = e^{-|b|^2/2}(-b)^{mN+2M} \frac{ \left\langle F
    \right\rangle_{N,M+m}}{I_N \sqrt{\tau}}.
\end{equation}

Therefore the zero mode tunnelling matrix
elements due to an impurity in the bulk can be computed for the two special
cases where $\chi = m$ given in terms of MC averages in (\ref{NtoA_qp3}) and
also for the case where $\chi = 1$ which has also been expressed in terms of MC
averages in the previous part of this section (\ref{V_chi1}).

\section{Results}

After showing the formulation for the microscopic computations, the results of
the simulations can now be discussed. They are showed in the later parts of this
section. To begin, details for the particular states and system sizes of MC
simulations are discussed.

\subsection{Simulation Details}

To run the MC simulations for the zero mode tunnelling matrix elements, the
filling fractions that were chosen for the simulations were $\nu=m^{-1}=1/3$ and
$\nu=m^{-1} = 1$. The free fermion case provides a good check for the numeric
methods used since for this state, all matrix elements of form (\ref{tme_qpChi})
can be calculated analytically.

When considering tunnelling across the bulk in a FQH device, there are two
particularly interesting cases. According to literature for Laughlin states the
most favourable form of tunnelling is for a single quasiparticle ($\chi =
1$). The other interesting case is for when an electron tunnels across the
bulk. In most other systems, as well as in the FQH system in the strong
backscattering regime, charge is usually transported by electrons. For these
reasons, MC simulations have been run for $\chi=1$ and $\chi = 3$. 

Expressions given by (\ref{V_chi1}) and (\ref{NtoA_qp3}) give the forms for the
tunnelling operators in terms of MC integrals for the tunnelling of a single
quasiparticle and three quasiparticles respectively. It is noted that in the
free fermion case where $\nu=1$, electrons are transported across the bulk
rather than Laughlin quasiparticles and for $\chi > 1$, all the matrix
elements in (\ref{tme_qpChi}) are zero. Therefore both methods presented in the
previous two sections to overcome the phase problem are equivalent to one
another in the free fermion case $\nu^{-1} = \chi = 1$. The results listed for
the free fermion case in the next section were obtained using method 1. Method 2
was used as a check for method 1 for which the same results were obtained. 

For the less trivial state $\nu=1/3$, MC calculations according to method 1 were
used to calculate the zero mode tunnelling matrix elements for a single
quasiparticle and method 2 was used for three quasiparticles/one electron
tunnelling. In method 1, for both $\nu=1$ and $\nu=1/3$ the cumulant expansion
(\ref{cumu_exp}) was computed up to the sixth cumulant. For method 2; there is
an additional value $d$ that appears in the integrals (\ref{F_average}) and
(\ref{IN_definition}) which was defined as some length away from the outer
boundary. An appropriate value to minimise systematic errors was found to be
$d=3$ magnetic lengths.

It is the system size dependence of the matrix elements given by
(\ref{tme_qpChi}) that is of interest and so multiple simulations where
performed for various values of $N$ ranging from 20 to 200 electrons. For all
values of $N$, the width of the system between the two edges was always kept
constant such that $R_O-R_I \sim 4$ units of magnetic length. This was achieved
by varying the value of $M$ accordingly with the number of electrons, $N$. The
only important statement about the placement of the impurity is that it was
equal distance from the inner and outer edge, i.e. $|b| = (R_O+R_I)/2$. Changing
the argument of $b$ has little physical effect on the tunnelling due to the
axial symmetry of the system. For simplicity these results were obtained by
choosing $b$ to be along the positive real axis.

\subsection{Tunnelling Results for $\nu=1$}

\begin{figure}[htb]
\begin{center}                        
\includegraphics[scale=1]{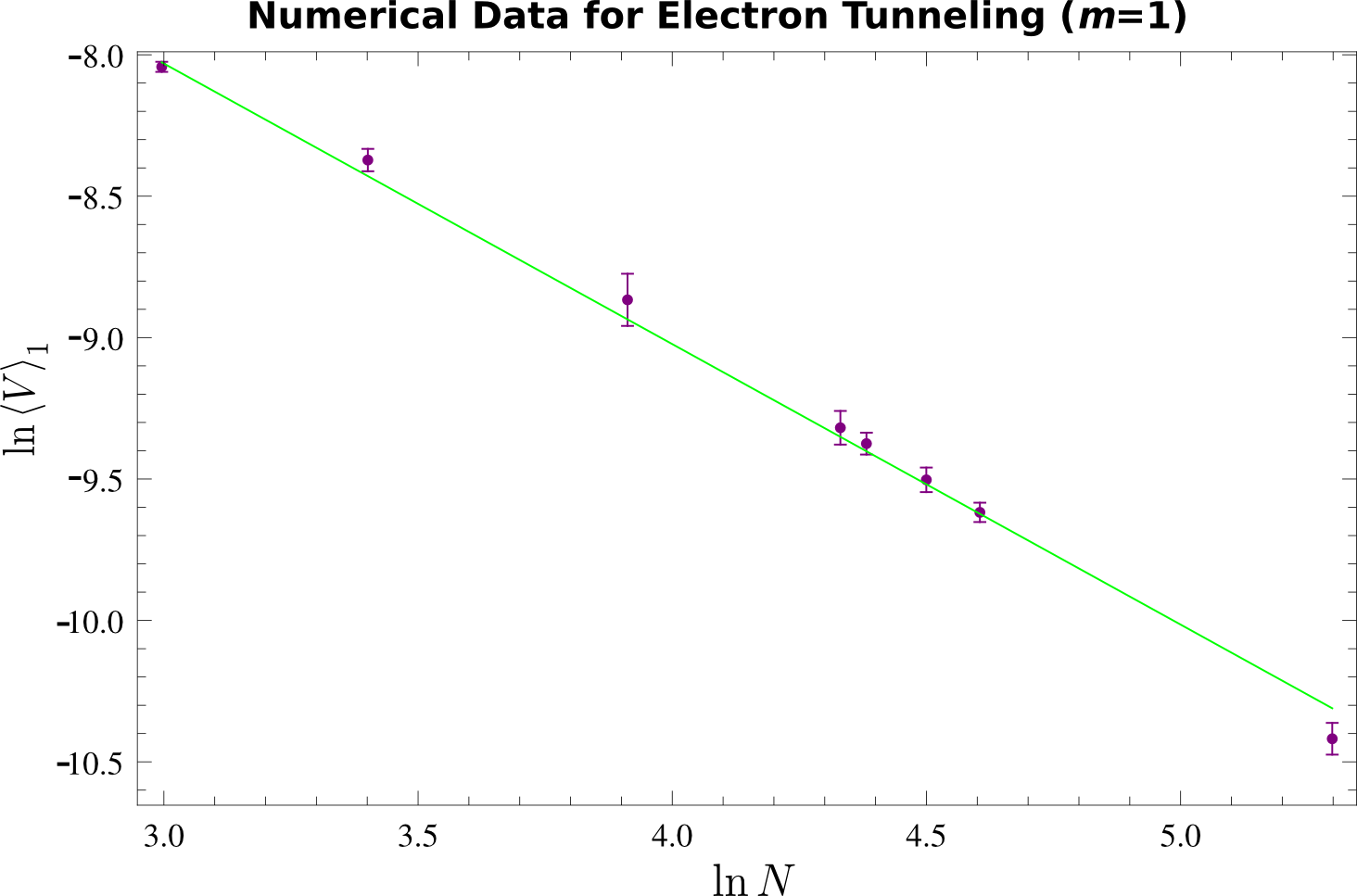}
\caption{Curve for the logarithm of the zero mode quasiparticle tunnelling
  operator versus the logarithm of $N$ as calculated using Monte Carlo for
  filling factor $\nu=m^{-1}=1$. The purple points are the plots from the Monte Carlo
  data whilst the green, linear curve is the line of best fit as shown in
  equation (\ref{tme_1qp_num_eq_1}).}
\label{tme_1qp_num_nu1}
\end{center}
\end{figure}

The zero mode tunnelling matrix elements in the free fermion case were
calculated microscopically according to (\ref{V_chi1}) where the averages were
computed using MC. The only non zero matrix element (excluding the trivial
$\chi=0$ case) is when a single electron is being transferred across the bulk
corresponding to $\chi=1$ in (\ref{V_chi1}). These results are presented
graphically on a log-log plot in Figure \ref{tme_1qp_num_nu1}.

The data set plotted on Figure \ref{tme_1qp_num_nu1} is fitted to straight a line
where the gradient of the line for $\nu = 1$ is given by,

\begin{equation} \label{tme_1qp_num_eq_1} \frac{\mathrm{d} \ln \left\langle V
    \right\rangle_1}{\mathrm{d} \ln N} = - 0.99 \pm 0.02 \quad \quad (\textrm{for} \quad \nu = 1)
\end{equation}

The tunnelling matrix elements obviously have a system size
dependence. Therefore when comparing these numerical results to the effective
theory results for the tunnelling operator, operator ordering defined in
(\ref{op_ordering}) must be imposed for a constant $t_p$. 

In the simulations, the impurity was placed at position $b$ along the real axis
and so in the effective theory calculation $\xi = 0$ and $R= |b|$. These
parameters allow us to drop the absolute value of the matrix elements since the
phase terms drop out anyway. Setting $p = \chi = 1$ in
(\ref{zm_tunnelling_me_el3}) gives

\begin{equation} \label{tme_1qp_SG_eq_1}
\frac{ \mathrm{d} \ln \left\langle A_1 \right\rangle}{\mathrm{d} \ln N} = -0.96 \quad \quad
(\textrm{for} \quad \nu = 1).
\end{equation}

Comparing (\ref{tme_1qp_SG_eq_1}) to (\ref{tme_1qp_num_eq_1}) shows that the
effective theory, when using non-normal ordered operators does match the
microscopic computations for the zero mode tunnelling matrix elements. The value
of the gradient of the plot in Figure \ref{tme_1qp_num_nu1} as predicted from
the effective theory is slightly out of the range of errors in the microscopic
calculation. Possible reasons for this will be discussed after the results have
been presented for the $\nu=1/3$ filling fraction.

\subsection{Tunnelling Results for $\nu=1/3$}

\begin{figure}[htb]
\begin{center}                        
\includegraphics[scale=1]{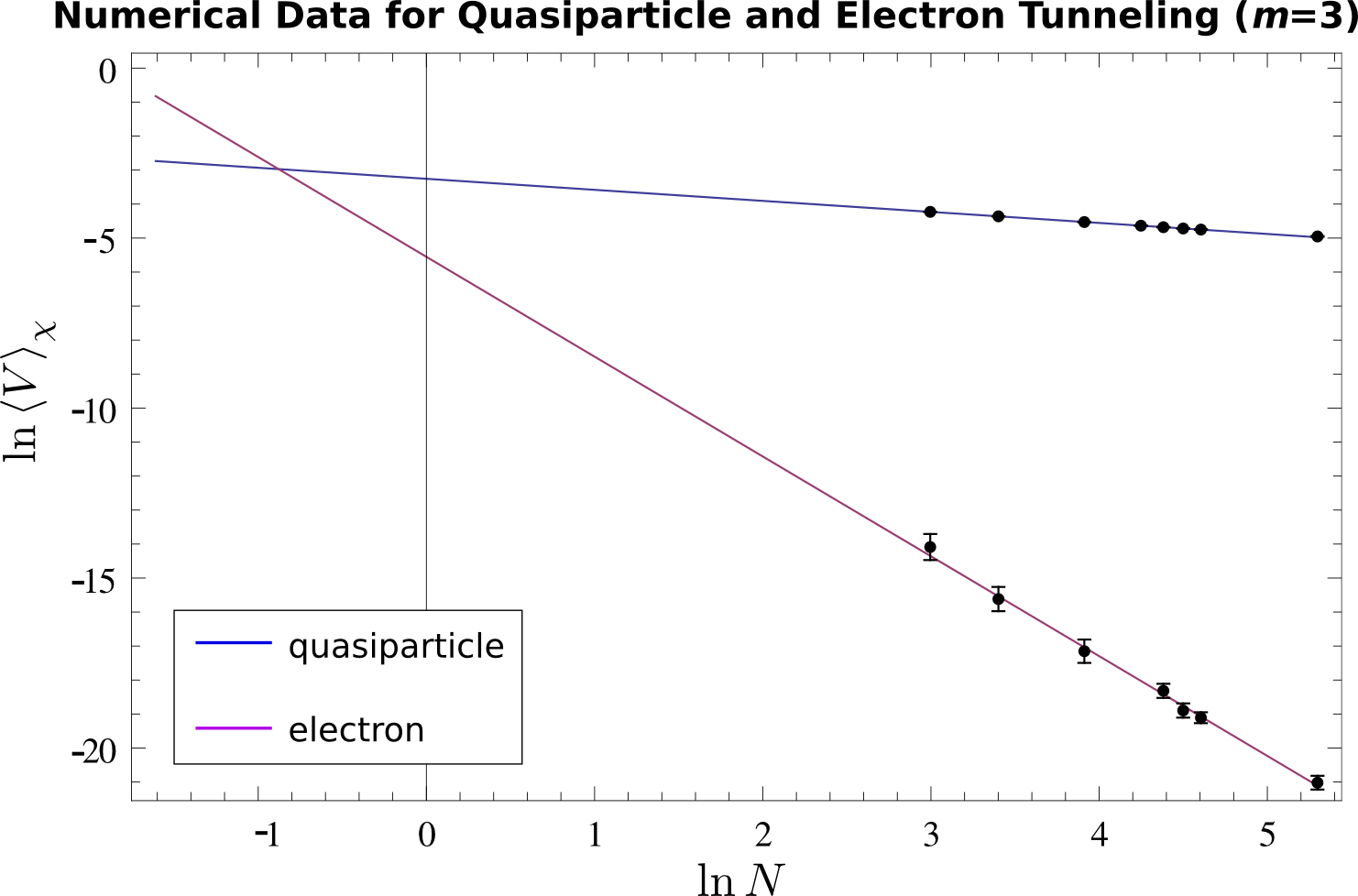}
\caption{Curve for the logarithm of the zero mode quasiparticle tunnelling
  operator versus the logarithm of $N$ as calculated using MC for filling factor
  $\nu=m^{-1}=1/3$. The black points are the plots from the MC data for $\chi=1$
  and $\chi=3$ whilst the blue and purple curves are lines of best fit shown in
  equations (\ref{tme_1qp_num_eq_3}) and (\ref{num_3qp_tme}) respectively.}
\label{m3_1qp_3qp}
\end{center}
\end{figure}

For the filling factor $\nu=1/3$ tunnelling matrix elements were computed for
both a quasiparticle ($\chi=1$) and an electron ($\chi=3$) tunnelling across the
bulk. Method 1 (\ref{V_chi1}) was used for the quasiparticle tunnelling case and
method 2 (\ref{NtoA_qp3}) for electron tunnelling. The
data from both MC simulations is presented on a log-log plot in Figure
\ref{m3_1qp_3qp}.

The gradient of linear curve fitted to the $\chi=1$ data set is;

\begin{equation} \label{tme_1qp_num_eq_3}
\frac{\mathrm{d}\ln \left\langle V \right\rangle_1}{\mathrm{d} \ln N} = - 0.325
\pm 0.001 \quad \quad
(\textrm{for} \quad \nu = 1/3).
\end{equation}

Following similar arguments to those given in the results section for $\nu=1$;
the effective theory prediction for the tunnelling matrix elements for a
quasiparticle transferred across the bulk is

\begin{equation} \label{tme_1qp_SG_eq_3}
\frac{\mathrm{d}\ln \left\langle A_1 \right\rangle} {\mathrm{d} \ln N} = - 0.333 \quad \quad
(\textrm{for} \quad \nu = 1/3).
\end{equation}

Comparing (\ref{tme_1qp_SG_eq_3}) to (\ref{tme_1qp_num_eq_3}); the effective
theory does predict the correct scaling behaviour for the zero mode tunnelling
matrix elements from the tunnelling Hamiltonian with non-normal ordered
operators. Similar to the case for $\chi=1$ however, the result is slightly out
of the error range of the microscopic calculations. 

A possible reason for this could be related to the fact that to over come the
phase problem in the original integral (\ref{tme_qpChi}) the cumulant expansion
was used. Obviously only a finite number of terms in the expansion could be
computed numerically and this will produce a systematic error in the value of
$\left\langle V \right\rangle_1$ for $\nu=1$ and $\nu=1/3$ that could account
for the differences in the predicted value from the effective theory and the
computed value from the microscopic theory. The cumulant expansion was
terminated when the magnitude of the next term in the expansion could no longer
be resolved due to the MC error. The number of terms kept varied for different
computations and was dependent on the number of electrons $N$ in the system. It
can be seen when comparing Figures \ref{m3_1qp_3qp} and
\ref{tme_1qp_num_nu1} that the errors for the simulation for $\nu=1/3$ are
smaller than those for $\nu=1$. This is due to the fact that the cumulant
expansion decayed much faster for $\nu=1/3$ than $\nu=1$ and so fewer terms in
the expansion for $\nu=1/3$ were kept.

The data for electron tunnelling ($\chi=3$) has also been fitted to a linear
curve in Figure \ref{m3_1qp_3qp}. The gradient of the line is given by;

\begin{equation} \label{num_3qp_tme}
\frac{\mathrm{d}\ln \left\langle V \right\rangle_3}{\mathrm{d} \ln N} = - 3.03\pm 0.09
\quad (\nu = 1/3). 
\end{equation}

From (\ref{zm_tunnelling_me_el3}), the effective theory prediction from the
non-normal ordered tunnelling Hamiltonian can be extracted for electron
tunnelling in the FQH state $\nu=1/3$.

\begin{equation} \label{SG_3qp_tme}
\frac{\mathrm{d}\ln \left\langle A_3 \right\rangle}{\mathrm{d} \ln N} = - 3.00 \quad (\nu = 1/3).
\end{equation}

For the case of an electron tunnelling across the bulk of a FQH device
the effective theory predictions for the scaling of the zero mode matrix
elements match the microscopic computations and are well within the error
range. In Figure \ref{m3_1qp_3qp} the curves describing the MC data set for a
quasiparticle and an electron tunnelling have been extrapolated such that the
point of intersection of the two curves can be seen. Interestingly, the point at
which the intersect occurs is when $N<1$ and therefore from the graph we see
that at all physically possible system sizes the electron tunnelling process is
always less relevant than the quasiparticle tunnelling process.

\section{Summary and Conclusions}

This work has investigated the zero mode tunnelling matrix elements due to an
impurity in the bulk which have been computed as a function of system size, $N$ and
then compared to the effective theory predictions for the tunnelling
operators. In the first section the effective theory predictions were
discussed. The quasiparticle operators from the Luttinger liquid theory of FQH
edge states were used to calculate the zero mode matrix elements of the
tunnelling operators $A_p$, where $p$ corresponded to the number of
quasiparticles tunnelling at the impurity. These matrix elements were calculate
using two types of ordering of quasiparticle operators. The first type was the
usual definition of normal ordering defined in (\ref{norm_ordering}) were it was
found that only the tunnelling parameter $t_p$ could contain system size
dependence. The second type of ordering considered was when the operators were
not normal ordered, as defined in (\ref{op_ordering}). These matrix elements did
show signs of system size dependence. To investigate which scaling of the
tunnelling parameters $t_p$ best describe the tunnelling events in a FQH system
a microscopic calculation was performed.

This microscopic calculation was based on the Laughlin states of the FQHE and was
the subject of section II. The microscopic formula that
describes the process for the tunnelling of $\chi$ particles due to the impurity
inserted in to the bulk is given by $\left\langle V \right\rangle_\chi$ in
(\ref{tme_qpChi}). The only known way of calculating such
integrals in (\ref{tme_qpChi}) was by using numerical methods. The MC method
was chosen for the computation of $\left\langle V \right\rangle_\chi$ though
to directly calculate this average would not be very efficient since it is an
average over a complex number which introduces a phase problem to the simulation.

Two methods were found to overcome this problem. Method 1. used MC to calculate
the cumulant expansion of a function related to $\left\langle V
\right\rangle_\chi$ and was suitable only for $\chi \leq 1$. Method 2. to overcome
the phase problem was applicable only for $\chi = \nu^{-1}$ in which case the
matrix elements in $\left\langle V \right\rangle_\chi$ could be written in terms
of absolute values of complex numbers thus avoiding any phase problems. 

Finally, the results of the MC calculations for $\left\langle V
\right\rangle_\chi$ were presented and compared to the effective theory
predictions of the tunnelling behaviour. It was found that the tunnelling
Hamiltonian predicted by the effective theory is an accurate representation of
the effects on the zero modes of the edges when an impurity is inserted into the
bulk of a $\nu=1/3$ Laughlin state. Thus the Luttinger liquid model at some
potential barrier has the same physics as a QPC inserted into a FQH device,
which is something that has been assumed to be correct but not rigorously tested
microscopically. When increasing the system sizes of the FQH device the electron
operator plays a less important role than that of the quasiparticle
operator. This can be seen for all system sizes and supports previous works
mentioned in the introduction when the the electron tunnelling term is dropped
in comparison to the quasiparticle tunnelling process. However, a match between
the numerical data and the predictions made from the tunnelling Hamiltonian
could only be obtained using a specific ordering for the operators. If one
wishes to make the tunnelling Hamiltonian normal ordered then there must be a
normalisation factor included in the definition of the quasiparticle and
electron operators. The behaviour of the tunnelling Hamiltonian using such
operators has also been shown to be well behaved physically, namely the operator
is local in the sense that its correlation function is system-size independent.

\bibliographystyle{apsrev}
\bibliography{my_bib}

\begin{thebibliography}{32}
\expandafter\ifx\csname natexlab\endcsname\relax\def\natexlab#1{#1}\fi
\expandafter\ifx\csname bibnamefont\endcsname\relax
  \def\bibnamefont#1{#1}\fi
\expandafter\ifx\csname bibfnamefont\endcsname\relax
  \def\bibfnamefont#1{#1}\fi
\expandafter\ifx\csname citenamefont\endcsname\relax
  \def\citenamefont#1{#1}\fi
\expandafter\ifx\csname url\endcsname\relax
  \def\url#1{\texttt{#1}}\fi
\expandafter\ifx\csname urlprefix\endcsname\relax\def\urlprefix{URL }\fi
\providecommand{\bibinfo}[2]{#2}
\providecommand{\eprint}[2][]{\url{#2}}

\bibitem[{\citenamefont{Tsui et~al.}(1982)\citenamefont{Tsui, Stormer, and
  Gossard}}]{fqhe_observation1}
\bibinfo{author}{\bibfnamefont{D.~C.} \bibnamefont{Tsui}},
  \bibinfo{author}{\bibfnamefont{H.~L.} \bibnamefont{Stormer}},
  \bibnamefont{and} \bibinfo{author}{\bibfnamefont{A.~C.}
  \bibnamefont{Gossard}}, \bibinfo{journal}{Phys. Rev. Lett.}
  \textbf{\bibinfo{volume}{48}}, \bibinfo{pages}{1559} (\bibinfo{year}{1982}).

\bibitem[{\citenamefont{Laughlin}(1983)}]{laughlinPA}
\bibinfo{author}{\bibfnamefont{R.~B.} \bibnamefont{Laughlin}},
  \bibinfo{journal}{Phys. Rev. Lett.} \textbf{\bibinfo{volume}{50}}
  (\bibinfo{year}{1983}),
  \urlprefix\url{http://link.aps.org/doi/10.1103/PhysRevLett.50.1395}.

\bibitem[{\citenamefont{Wen}(1990)}]{wenChiLL}
\bibinfo{author}{\bibfnamefont{X.~G.} \bibnamefont{Wen}},
  \bibinfo{journal}{Phys. Rev. B} \textbf{\bibinfo{volume}{41}}
  (\bibinfo{year}{1990}),
  \urlprefix\url{http://link.aps.org/doi/10.1103/PhysRevB.41.12838}.

\bibitem[{\citenamefont{Chang}(2003)}]{RevModPhys.75.1449}
\bibinfo{author}{\bibfnamefont{A.~M.} \bibnamefont{Chang}},
  \bibinfo{journal}{Rev. Mod. Phys.} \textbf{\bibinfo{volume}{75}},
  \bibinfo{pages}{1449} (\bibinfo{year}{2003}),
  \urlprefix\url{http://link.aps.org/doi/10.1103/RevModPhys.75.1449}.

\bibitem[{\citenamefont{Kane and
  Fisher}(1992{\natexlab{a}})}]{kane1992transmission}
\bibinfo{author}{\bibfnamefont{C.}~\bibnamefont{Kane}} \bibnamefont{and}
  \bibinfo{author}{\bibfnamefont{M.~P.} \bibnamefont{Fisher}},
  \bibinfo{journal}{Physical Review B} \textbf{\bibinfo{volume}{46}},
  \bibinfo{pages}{15233} (\bibinfo{year}{1992}{\natexlab{a}}).

\bibitem[{\citenamefont{Kane and
  Fisher}(1992{\natexlab{b}})}]{PhysRevB.46.7268}
\bibinfo{author}{\bibfnamefont{C.~L.} \bibnamefont{Kane}} \bibnamefont{and}
  \bibinfo{author}{\bibfnamefont{M.~P.~A.} \bibnamefont{Fisher}},
  \bibinfo{journal}{Phys. Rev. B} \textbf{\bibinfo{volume}{46}},
  \bibinfo{pages}{7268} (\bibinfo{year}{1992}{\natexlab{b}}),
  \urlprefix\url{http://link.aps.org/doi/10.1103/PhysRevB.46.7268}.

\bibitem[{\citenamefont{Kane and
  Fisher}(1992{\natexlab{c}})}]{PhysRevLett.68.1220}
\bibinfo{author}{\bibfnamefont{C.~L.} \bibnamefont{Kane}} \bibnamefont{and}
  \bibinfo{author}{\bibfnamefont{M.~P.~A.} \bibnamefont{Fisher}},
  \bibinfo{journal}{Phys. Rev. Lett.} \textbf{\bibinfo{volume}{68}},
  \bibinfo{pages}{1220} (\bibinfo{year}{1992}{\natexlab{c}}),
  \urlprefix\url{http://link.aps.org/doi/10.1103/PhysRevLett.68.1220}.

\bibitem[{\citenamefont{Furusaki and Nagaosa}(1993)}]{PhysRevB.47.4631}
\bibinfo{author}{\bibfnamefont{A.}~\bibnamefont{Furusaki}} \bibnamefont{and}
  \bibinfo{author}{\bibfnamefont{N.}~\bibnamefont{Nagaosa}},
  \bibinfo{journal}{Phys. Rev. B} \textbf{\bibinfo{volume}{47}},
  \bibinfo{pages}{4631} (\bibinfo{year}{1993}),
  \urlprefix\url{http://link.aps.org/doi/10.1103/PhysRevB.47.4631}.

\bibitem[{\citenamefont{Moon et~al.}(1993)\citenamefont{Moon, Yi, Kane, Girvin,
  and Fisher}}]{PhysRevLett.71.4381}
\bibinfo{author}{\bibfnamefont{K.}~\bibnamefont{Moon}},
  \bibinfo{author}{\bibfnamefont{H.}~\bibnamefont{Yi}},
  \bibinfo{author}{\bibfnamefont{C.~L.} \bibnamefont{Kane}},
  \bibinfo{author}{\bibfnamefont{S.~M.} \bibnamefont{Girvin}},
  \bibnamefont{and} \bibinfo{author}{\bibfnamefont{M.~P.~A.}
  \bibnamefont{Fisher}}, \bibinfo{journal}{Phys. Rev. Lett.}
  \textbf{\bibinfo{volume}{71}}, \bibinfo{pages}{4381} (\bibinfo{year}{1993}),
  \urlprefix\url{http://link.aps.org/doi/10.1103/PhysRevLett.71.4381}.

\bibitem[{\citenamefont{Gogolin et~al.}(2004)\citenamefont{Gogolin, Nersesyan,
  and Tsvelik}}]{gogolin2004bosonization}
\bibinfo{author}{\bibfnamefont{A.}~\bibnamefont{Gogolin}},
  \bibinfo{author}{\bibfnamefont{A.}~\bibnamefont{Nersesyan}},
  \bibnamefont{and} \bibinfo{author}{\bibfnamefont{A.}~\bibnamefont{Tsvelik}},
  \emph{\bibinfo{title}{Bosonization and Strongly Correlated Systems}}
  (\bibinfo{publisher}{Cambridge University Press}, \bibinfo{year}{2004}), ISBN
  \bibinfo{isbn}{9780521617192},
  \urlprefix\url{http://books.google.co.uk/books?id=BZDfFIpCoaAC}.

\bibitem[{\citenamefont{Milliken et~al.}(1996)\citenamefont{Milliken, Umbach,
  and Webb}}]{Milliken1996309}
\bibinfo{author}{\bibfnamefont{F.}~\bibnamefont{Milliken}},
  \bibinfo{author}{\bibfnamefont{C.}~\bibnamefont{Umbach}}, \bibnamefont{and}
  \bibinfo{author}{\bibfnamefont{R.}~\bibnamefont{Webb}},
  \bibinfo{journal}{Solid State Communications} \textbf{\bibinfo{volume}{97}},
  \bibinfo{pages}{309 } (\bibinfo{year}{1996}), ISSN \bibinfo{issn}{0038-1098},
  \urlprefix\url{http://www.sciencedirect.com/science/article/pii/003810989500%
1816}.

\bibitem[{\citenamefont{Roddaro et~al.}(2004)\citenamefont{Roddaro, Pellegrini,
  Beltram, Biasiol, and Sorba}}]{PhysRevLett.93.046801}
\bibinfo{author}{\bibfnamefont{S.}~\bibnamefont{Roddaro}},
  \bibinfo{author}{\bibfnamefont{V.}~\bibnamefont{Pellegrini}},
  \bibinfo{author}{\bibfnamefont{F.}~\bibnamefont{Beltram}},
  \bibinfo{author}{\bibfnamefont{G.}~\bibnamefont{Biasiol}}, \bibnamefont{and}
  \bibinfo{author}{\bibfnamefont{L.}~\bibnamefont{Sorba}},
  \bibinfo{journal}{Phys. Rev. Lett.} \textbf{\bibinfo{volume}{93}},
  \bibinfo{pages}{046801} (\bibinfo{year}{2004}),
  \urlprefix\url{http://link.aps.org/doi/10.1103/PhysRevLett.93.046801}.

\bibitem[{\citenamefont{Chang et~al.}(1996)\citenamefont{Chang, Pfeiffer, and
  West}}]{PhysRevLett.77.2538}
\bibinfo{author}{\bibfnamefont{A.~M.} \bibnamefont{Chang}},
  \bibinfo{author}{\bibfnamefont{L.~N.} \bibnamefont{Pfeiffer}},
  \bibnamefont{and} \bibinfo{author}{\bibfnamefont{K.~W.} \bibnamefont{West}},
  \bibinfo{journal}{Phys. Rev. Lett.} \textbf{\bibinfo{volume}{77}},
  \bibinfo{pages}{2538} (\bibinfo{year}{1996}),
  \urlprefix\url{http://link.aps.org/doi/10.1103/PhysRevLett.77.2538}.

\bibitem[{\citenamefont{Kane and Fisher}(1994)}]{PhysRevLett.72.724}
\bibinfo{author}{\bibfnamefont{C.~L.} \bibnamefont{Kane}} \bibnamefont{and}
  \bibinfo{author}{\bibfnamefont{M.~P.~A.} \bibnamefont{Fisher}},
  \bibinfo{journal}{Phys. Rev. Lett.} \textbf{\bibinfo{volume}{72}},
  \bibinfo{pages}{724} (\bibinfo{year}{1994}),
  \urlprefix\url{http://link.aps.org/doi/10.1103/PhysRevLett.72.724}.

\bibitem[{\citenamefont{Saminadayar et~al.}(1997)\citenamefont{Saminadayar,
  Glattli, Jin, and Etienne}}]{shotnoise_support2}
\bibinfo{author}{\bibfnamefont{L.}~\bibnamefont{Saminadayar}},
  \bibinfo{author}{\bibfnamefont{D.~C.} \bibnamefont{Glattli}},
  \bibinfo{author}{\bibfnamefont{Y.}~\bibnamefont{Jin}}, \bibnamefont{and}
  \bibinfo{author}{\bibfnamefont{B.}~\bibnamefont{Etienne}},
  \bibinfo{journal}{Phys. Rev. Lett.} \textbf{\bibinfo{volume}{79}},
  \bibinfo{pages}{2526} (\bibinfo{year}{1997}),
  \urlprefix\url{http://link.aps.org/doi/10.1103/PhysRevLett.79.2526}.

\bibitem[{\citenamefont{dePicciotto et~al.}(1997)\citenamefont{dePicciotto,
  Reznikov, Heiblum, Umansky, Bunin, and Mahalu}}]{shotnoise_support1}
\bibinfo{author}{\bibfnamefont{R.}~\bibnamefont{dePicciotto}},
  \bibinfo{author}{\bibfnamefont{M.}~\bibnamefont{Reznikov}},
  \bibinfo{author}{\bibfnamefont{M.}~\bibnamefont{Heiblum}},
  \bibinfo{author}{\bibfnamefont{V.}~\bibnamefont{Umansky}},
  \bibinfo{author}{\bibfnamefont{G.}~\bibnamefont{Bunin}}, \bibnamefont{and}
  \bibinfo{author}{\bibfnamefont{D.}~\bibnamefont{Mahalu}},
  \bibinfo{journal}{NATURE} \textbf{\bibinfo{volume}{389}},
  \bibinfo{pages}{162} (\bibinfo{year}{1997}), ISSN \bibinfo{issn}{0028-0836}.

\bibitem[{\citenamefont{Dolev et~al.}(2010)\citenamefont{Dolev, Gross, Chung,
  Heiblum, Umansky, and Mahalu}}]{shotnoise_question}
\bibinfo{author}{\bibfnamefont{M.}~\bibnamefont{Dolev}},
  \bibinfo{author}{\bibfnamefont{Y.}~\bibnamefont{Gross}},
  \bibinfo{author}{\bibfnamefont{Y.~C.} \bibnamefont{Chung}},
  \bibinfo{author}{\bibfnamefont{M.}~\bibnamefont{Heiblum}},
  \bibinfo{author}{\bibfnamefont{V.}~\bibnamefont{Umansky}}, \bibnamefont{and}
  \bibinfo{author}{\bibfnamefont{D.}~\bibnamefont{Mahalu}},
  \bibinfo{journal}{Phys. Rev. B} \textbf{\bibinfo{volume}{81}},
  \bibinfo{pages}{161303} (\bibinfo{year}{2010}).

\bibitem[{\citenamefont{Baer et~al.}(2014)\citenamefont{Baer, R\"ossler, Ihn,
  Ensslin, Reichl, and Wegscheider}}]{PhysRevB.90.075403}
\bibinfo{author}{\bibfnamefont{S.}~\bibnamefont{Baer}},
  \bibinfo{author}{\bibfnamefont{C.}~\bibnamefont{R\"ossler}},
  \bibinfo{author}{\bibfnamefont{T.}~\bibnamefont{Ihn}},
  \bibinfo{author}{\bibfnamefont{K.}~\bibnamefont{Ensslin}},
  \bibinfo{author}{\bibfnamefont{C.}~\bibnamefont{Reichl}}, \bibnamefont{and}
  \bibinfo{author}{\bibfnamefont{W.}~\bibnamefont{Wegscheider}},
  \bibinfo{journal}{Phys. Rev. B} \textbf{\bibinfo{volume}{90}},
  \bibinfo{pages}{075403} (\bibinfo{year}{2014}),
  \urlprefix\url{http://link.aps.org/doi/10.1103/PhysRevB.90.075403}.

\bibitem[{\citenamefont{Yang and Feldman}(2013)}]{PhysRevB.88.085317}
\bibinfo{author}{\bibfnamefont{G.}~\bibnamefont{Yang}} \bibnamefont{and}
  \bibinfo{author}{\bibfnamefont{D.~E.} \bibnamefont{Feldman}},
  \bibinfo{journal}{Phys. Rev. B} \textbf{\bibinfo{volume}{88}},
  \bibinfo{pages}{085317} (\bibinfo{year}{2013}),
  \urlprefix\url{http://link.aps.org/doi/10.1103/PhysRevB.88.085317}.

\bibitem[{\citenamefont{Fendley et~al.}(1995)\citenamefont{Fendley, Ludwig, and
  Saleur}}]{PhysRevLett.74.3005}
\bibinfo{author}{\bibfnamefont{P.}~\bibnamefont{Fendley}},
  \bibinfo{author}{\bibfnamefont{A.~W.~W.} \bibnamefont{Ludwig}},
  \bibnamefont{and} \bibinfo{author}{\bibfnamefont{H.}~\bibnamefont{Saleur}},
  \bibinfo{journal}{Phys. Rev. Lett.} \textbf{\bibinfo{volume}{74}},
  \bibinfo{pages}{3005} (\bibinfo{year}{1995}),
  \urlprefix\url{http://link.aps.org/doi/10.1103/PhysRevLett.74.3005}.

\bibitem[{\citenamefont{Guyon et~al.}(2002)\citenamefont{Guyon, Devillard,
  Martin, and Safi}}]{PhysRevB.65.153304}
\bibinfo{author}{\bibfnamefont{R.}~\bibnamefont{Guyon}},
  \bibinfo{author}{\bibfnamefont{P.}~\bibnamefont{Devillard}},
  \bibinfo{author}{\bibfnamefont{T.}~\bibnamefont{Martin}}, \bibnamefont{and}
  \bibinfo{author}{\bibfnamefont{I.}~\bibnamefont{Safi}},
  \bibinfo{journal}{Phys. Rev. B} \textbf{\bibinfo{volume}{65}},
  \bibinfo{pages}{153304} (\bibinfo{year}{2002}),
  \urlprefix\url{http://link.aps.org/doi/10.1103/PhysRevB.65.153304}.

\bibitem[{\citenamefont{Kane}(2003)}]{PhysRevLett.90.226802}
\bibinfo{author}{\bibfnamefont{C.~L.} \bibnamefont{Kane}},
  \bibinfo{journal}{Phys. Rev. Lett.} \textbf{\bibinfo{volume}{90}},
  \bibinfo{pages}{226802} (\bibinfo{year}{2003}),
  \urlprefix\url{http://link.aps.org/doi/10.1103/PhysRevLett.90.226802}.

\bibitem[{\citenamefont{Safi et~al.}(2001)\citenamefont{Safi, Devillard, and
  Martin}}]{PhysRevLett.86.4628}
\bibinfo{author}{\bibfnamefont{I.}~\bibnamefont{Safi}},
  \bibinfo{author}{\bibfnamefont{P.}~\bibnamefont{Devillard}},
  \bibnamefont{and} \bibinfo{author}{\bibfnamefont{T.}~\bibnamefont{Martin}},
  \bibinfo{journal}{Phys. Rev. Lett.} \textbf{\bibinfo{volume}{86}},
  \bibinfo{pages}{4628} (\bibinfo{year}{2001}),
  \urlprefix\url{http://link.aps.org/doi/10.1103/PhysRevLett.86.4628}.

\bibitem[{\citenamefont{Law et~al.}(2006)\citenamefont{Law, Feldman, and
  Gefen}}]{PhysRevB.74.045319}
\bibinfo{author}{\bibfnamefont{K.~T.} \bibnamefont{Law}},
  \bibinfo{author}{\bibfnamefont{D.~E.} \bibnamefont{Feldman}},
  \bibnamefont{and} \bibinfo{author}{\bibfnamefont{Y.}~\bibnamefont{Gefen}},
  \bibinfo{journal}{Phys. Rev. B} \textbf{\bibinfo{volume}{74}},
  \bibinfo{pages}{045319} (\bibinfo{year}{2006}),
  \urlprefix\url{http://link.aps.org/doi/10.1103/PhysRevB.74.045319}.

\bibitem[{\citenamefont{Vishveshwara}(2003)}]{PhysRevLett.91.196803}
\bibinfo{author}{\bibfnamefont{S.}~\bibnamefont{Vishveshwara}},
  \bibinfo{journal}{Phys. Rev. Lett.} \textbf{\bibinfo{volume}{91}},
  \bibinfo{pages}{196803} (\bibinfo{year}{2003}),
  \urlprefix\url{http://link.aps.org/doi/10.1103/PhysRevLett.91.196803}.

\bibitem[{\citenamefont{Nayak et~al.}(1999)\citenamefont{Nayak, Fisher, Ludwig,
  and Lin}}]{PhysRevB.59.15694}
\bibinfo{author}{\bibfnamefont{C.}~\bibnamefont{Nayak}},
  \bibinfo{author}{\bibfnamefont{M.~P.~A.} \bibnamefont{Fisher}},
  \bibinfo{author}{\bibfnamefont{A.~W.~W.} \bibnamefont{Ludwig}},
  \bibnamefont{and} \bibinfo{author}{\bibfnamefont{H.~H.} \bibnamefont{Lin}},
  \bibinfo{journal}{Phys. Rev. B} \textbf{\bibinfo{volume}{59}},
  \bibinfo{pages}{15694} (\bibinfo{year}{1999}),
  \urlprefix\url{http://link.aps.org/doi/10.1103/PhysRevB.59.15694}.

\bibitem[{\citenamefont{Ponomarenko and Averin}(2004)}]{PhysRevB.70.195316}
\bibinfo{author}{\bibfnamefont{V.~V.} \bibnamefont{Ponomarenko}}
  \bibnamefont{and} \bibinfo{author}{\bibfnamefont{D.~V.}
  \bibnamefont{Averin}}, \bibinfo{journal}{Phys. Rev. B}
  \textbf{\bibinfo{volume}{70}}, \bibinfo{pages}{195316}
  (\bibinfo{year}{2004}),
  \urlprefix\url{http://link.aps.org/doi/10.1103/PhysRevB.70.195316}.

\bibitem[{\citenamefont{Jonckheere et~al.}(2005)\citenamefont{Jonckheere,
  Devillard, Cr\'epieux, and Martin}}]{PhysRevB.72.201305}
\bibinfo{author}{\bibfnamefont{T.}~\bibnamefont{Jonckheere}},
  \bibinfo{author}{\bibfnamefont{P.}~\bibnamefont{Devillard}},
  \bibinfo{author}{\bibfnamefont{A.}~\bibnamefont{Cr\'epieux}},
  \bibnamefont{and} \bibinfo{author}{\bibfnamefont{T.}~\bibnamefont{Martin}},
  \bibinfo{journal}{Phys. Rev. B} \textbf{\bibinfo{volume}{72}},
  \bibinfo{pages}{201305} (\bibinfo{year}{2005}),
  \urlprefix\url{http://link.aps.org/doi/10.1103/PhysRevB.72.201305}.

\bibitem[{\citenamefont{Haldane}(1983)}]{PhysRevLett.51.605}
\bibinfo{author}{\bibfnamefont{F.~D.~M.} \bibnamefont{Haldane}},
  \bibinfo{journal}{Phys. Rev. Lett.} \textbf{\bibinfo{volume}{51}},
  \bibinfo{pages}{605} (\bibinfo{year}{1983}),
  \urlprefix\url{http://link.aps.org/doi/10.1103/PhysRevLett.51.605}.

\bibitem[{\citenamefont{A.~Boyarsky}(2004)}]{vadim1}
\bibinfo{author}{\bibfnamefont{O.~R.} \bibnamefont{A.~Boyarsky},
  \bibfnamefont{Vadim V.~Cheianov}}, \bibinfo{journal}{arXiv:cond-mat/0402562v2
  [cond-mat.mes-hall]}  (\bibinfo{year}{2004}).

\bibitem[{\citenamefont{Ivan P.~Levkivskyi}(2010)}]{LFS_ll_annulus}
\bibinfo{author}{\bibfnamefont{E.~V.~S.} \bibnamefont{Ivan P.~Levkivskyi},
  \bibfnamefont{Juerg~Froehlich}}, \bibinfo{journal}{arXiv:1005.5703v1
  [cond-mat.mes-hall]}  (\bibinfo{year}{2010}).

\bibitem[{\citenamefont{Metropolis et~al.}(1953)\citenamefont{Metropolis,
  Rosenbluth, Rosenbluth, Teller, and Teller}}]{metropolisAlg}
\bibinfo{author}{\bibfnamefont{N.}~\bibnamefont{Metropolis}},
  \bibinfo{author}{\bibfnamefont{A.~W.} \bibnamefont{Rosenbluth}},
  \bibinfo{author}{\bibfnamefont{M.~N.} \bibnamefont{Rosenbluth}},
  \bibinfo{author}{\bibfnamefont{A.~H.} \bibnamefont{Teller}},
  \bibnamefont{and} \bibinfo{author}{\bibfnamefont{E.}~\bibnamefont{Teller}},
  \bibinfo{journal}{The Journal of Chemical Physics}
  \textbf{\bibinfo{volume}{21}}, \bibinfo{pages}{1087} (\bibinfo{year}{1953}),
  \urlprefix\url{http://scitation.aip.org/content/aip/journal/jcp/21/6/10.1063%
/1.1699114}.

\end{thebibliography}
\end{document}